\documentclass{article}

\usepackage{microtype}
\usepackage{graphicx}
\usepackage{subfigure}
\usepackage{booktabs} 
\usepackage{amsmath}
\allowdisplaybreaks

\usepackage{hyperref}

\newcommand{\jaxcosmo}{\texttt{jax-cosmo}}

\usepackage[accepted]{icml2025}

\usepackage{amsmath}
\usepackage{amssymb}
\usepackage{mathtools}
\usepackage{amsthm}

\usepackage[capitalize,noabbrev]{cleveref}

\theoremstyle{plain}

\theoremstyle{definition}

\theoremstyle{remark}

\usepackage[textsize=tiny]{todonotes}

\icmltitlerunning{Fisher Score Matching}

\begin{document}

\twocolumn[
\icmltitle{Fisher Score Matching for Simulation-Based Forecasting and Inference}



\icmlsetsymbol{equal}{*}

\begin{icmlauthorlist}
\icmlauthor{Ce Sui}{tsinghua,jhupha}
\icmlauthor{Shivam Pandey}{jhupha}
\icmlauthor{Benjamin D. Wandelt}{jhupha,jhuams}
\end{icmlauthorlist}

\icmlaffiliation{tsinghua}{Department of Astronomy, Tsinghua University, Beijing 100084, China.}
\icmlaffiliation{jhupha}{Department of Physics and Astronomy, Johns Hopkins University, Baltimore, MD 21218, USA}
\icmlaffiliation{jhuams}{Department of Applied Mathematics and Statistics, Johns Hopkins University, Baltimore, MD 21218, USA}

\icmlcorrespondingauthor{Ce Sui}{suic20@mails.tsinghua.edu.cn}

\icmlkeywords{Machine Learning, Cosmology, Simulation-based Inference}

\vskip 0.3in
]



\printAffiliationsAndNotice{}  

\begin{abstract}
We propose a method for estimating the Fisher score—the gradient of the log-likelihood with respect to model parameters—using score matching. By introducing a latent parameter model, we show that the Fisher score can be learned by training a neural network to predict latent scores via a mean squared error loss. We validate our approach on a toy linear Gaussian model and a cosmological example using a differentiable simulator. In both cases, the learned scores closely match ground truth for  plausible data-parameter pairs. This method extends the ability to perform Fisher forecasts, and gradient-based Bayesian inference to simulation models, even when they are not differentiable; it therefore has broad potential for advancing cosmological analyses.
\end{abstract}

\section{Introduction}
\label{intro}
Statistical inference is a core component of modern astronomical and cosmological research. A typical inference pipeline involves several key stages: identifying informative observables, constructing  summary statistics, and performing parameter inference. A wide array of tools has been developed to address each of these tasks.

Decades of research on cosmological structure have centered on defining informative summaries and observables, principally amongst them the power spectrum.  New surveys, such as those focusing on galaxy clustering, 21cm intensity mapping, and weak lensing \cite{DESIY3,Euclid,SKA} motivate more sophisticated summaries  \cite{beyond2pt,hybrid,IMNN,scatteringtransform}. These include both hand-crafted features and machine learning-based methods, aimed at extracting the maximal amount of information from increasingly complex data. To forecast constraints based on these observables and summaries, the Fisher information matrix has been widely used after being introduced to cosmology by \citet{Tegmark_fisher}.  

Given observational data summary statistics and a physical model, the next step is parameter inference. Most current cosmological analyses use Bayesian inference (\textit{e.g.},~\citet{DESIY3,Plank18}). Traditionally, this involves assuming an explicit form for the likelihood and using MCMC algorithms for posterior sampling. Numerous MCMC packages have been developed to support this \cite{Cobaya,MontePython}. More recently, simulation-based inference (SBI) \cite{sbi_review,pydelfi} has allowed approximating the likelihood implicitly using simulations. SBI methods can  be more flexible and efficient than standard MCMC, and many recent tools implement SBI techniques to address a wide range of cosmological problems \cite{ltu-ili,sbi_package,xiaosheng_21cm_sbi,simbig}.

All of these steps—summary construction, Fisher forecasting, and inference—can be greatly simplified if we have access to the Fisher score function: the gradient of the log-likelihood with respect to the model parameters. The Fisher score function itself is a sufficient statistic, since it encodes the likelihood surface up to a constant factor. When evaluated at a fixed, fiducial parameter value, it serves as a locally sufficient statistic \cite{MOPED,fisherscoresummary} and enables straightforward computation of the Fisher matrix. It serves to construct unbiased, minimum-variance estimators that saturate the Cramér-Rao information inequality, and to compute the maximum likelihood estimator. Furthermore, it opens the door to using more efficient gradient-based approaches such as Hamiltonian Monte Carlo (HMC) samplers; or to frequentist analysis avoiding prior choice, such as maximum likelihood estimators.

This surprising utility of the Fisher score has motivated recent efforts to re-implement cosmological simulations using differentiable programming frameworks like JAX\footnote{\url{http://github.com/jax-ml/jax}} \citep{jax2018github} that enable automatic differentiation throughout the simulation pipeline and make Fisher score computation feasible \cite{pmwd,jaxcosmo}. In parallel, score matching—a technique used to train neural networks to approximate not the Fisher score, but the probability score $\nabla_{x}\log P(x)$—has seen widespread success in generative modeling, most notably in diffusion models \cite{NCSM,sde_diffusion}. Similar techniques have also been adapted to the context of SBI and cosmological inference tasks \cite{diffusion_sbi,diffusion-cosmo,diffusion-cosmo-2}.

Here, we propose a new score matching approach for learning the Fisher score directly from simulations. We demonstrate that the learned scores are accurate for data-parameter pairs relevant during inference and can be used for both Fisher analysis and Bayesian inference. Section~\ref{sec:method} describes our method, Section~\ref{sec:ex} presents experimental results, and Section~\ref{sec:conclusion} concludes with a discussion of future directions.

\section{Method}
\label{sec:method}

Given a probabilistic model $P(x \mid \theta)$ describing a physical system, a key quantity of interest is the Fisher score, defined as:  
\begin{equation}
    s(x, \theta) = \nabla_{\theta} \log P(x \mid \theta).
\end{equation}
However, in many cases the likelihood $P(x \mid \theta)$ is only implicitly defined through a simulator, without an analytical form or differentiable implementation, making direct computation of the Fisher score intractable. We propose a method to estimate the Fisher score using score matching.

\subsection{Fisher Score Matching}
We assume the existence of a latent variable $\theta^*$ such that the following Markov chain holds: $\theta \rightarrow \theta^* \rightarrow x$. That is,
\[
P(x \mid \theta) = \int P(x \mid \theta^*) P(\theta^* \mid \theta) \, d\theta^*,
\]
where $P(\theta^* \mid \theta)$ defines the latent model. Then we can re-write the Fisher score of the full model as an expectation of the latent model’s score as follows: 
\begin{align}
    \nabla_{\theta}\log P(x|\theta)&= \frac{\nabla_{\theta} P(x|\theta)}{P(x|\theta)}\notag\\
    &=\frac{\nabla_{\theta}\int P(x|\theta^*,\theta)P(\theta^*|\theta)d\theta^*}{P(x|\theta)}\notag\\
    &=\frac{\int P(x|\theta^*)\nabla_{\theta}P(\theta^*|\theta)d\theta^*}{P(x|\theta)}\notag\\
    &=\int\frac{ P(x|\theta^*)P(\theta^*|\theta)}{P(x|\theta)}\nabla_{\theta}\log P(\theta^*|\theta)d\theta^*\notag\\
    &=\int P(\theta^*|x,\theta)\nabla_{\theta}\log P(\theta^*|\theta)d\theta^*\notag\\
    &=E_{P(\theta^*|x,\theta)}[\nabla_{\theta}\log P(\theta^*|\theta)].
\end{align}
Since posterior means can be estimated by minimizing mean squared error (see Appendix~\ref{app:bls}), we can train a score estimator $s(x, \theta)$ by minimizing the  loss
\begin{equation}
    L = E_{p(x,\theta, \theta^*)}[(s(x,\theta)-\nabla_{\theta}\log P(\theta^*|\theta))^2]
    \label{eq:FisherScoreLoss}
\end{equation}
Training proceeds as follows: we sample from the joint distribution \( p(x, \theta, \theta^*) \) using forward simulations. For each triplet \( (x, \theta, \theta^*) \), the model is asked to predict the latent score \( \nabla_{\theta} \log p(\theta^* \mid \theta) \) given the input \( (x, \theta) \). Once training converges, the estimator \( s(x, \theta) \) provides a deterministic approximation of the true Fisher score \( \nabla_{\theta} \log p(x \mid \theta) \). Since the model is essentially performing a regression task, we expect its scalability with input dimension to follow typical patterns observed in regression problems, depending on the architecture used to learn the score.

\subsection{Decomposable and Indecomposable Models}
If we can identify a latent variable such that the Markov chain $\theta \rightarrow \theta^* \rightarrow x$ holds and the gradient $\nabla_{\theta} \log P(\theta^* \mid \theta)$ is tractable, this approach can be applied directly, a fact  also exploited by \citet{Brehmer_2020}.

A simple example is the linear Gaussian model $x \mid \theta \sim \mathcal{N}(\theta, \Sigma_n)$. We can decompose this into two conditional models
\begin{equation}
\label{gaussian_model}
    \theta^* \mid \theta \sim \mathcal{N}(\theta, \Sigma_{\theta^*}), \quad x \mid \theta^* \sim \mathcal{N}(\theta^*, \Sigma_n - \Sigma_{\theta^*}),
\end{equation}

where both covariance matrices are positive definite. The Fisher score of the full model can then be learned using only the score from the latent model.

In this case, the choice of decomposition does not affect the result—the training objective will still converge to the true Fisher score of the full model.

To address more complex settings where no natural latent variable is available, we  introduce an auxiliary variable \( \tilde{\theta} \) and define a latent model \( P(\theta \mid \tilde{\theta}) \), forming the Markov chain \( \tilde{\theta} \rightarrow \theta \rightarrow x \). We simulate samples using
\[
P(\theta, x \mid \tilde{\theta}) = P(x \mid \theta) P(\theta \mid \tilde{\theta}),
\]
and apply the same score matching technique to estimate $\nabla_{\tilde{\theta}} \log P(x \mid \tilde{\theta})$. This approximation becomes accurate when $P(\theta \mid \tilde{\theta})$ is sharply peaked (i.e., close to a delta function). A natural choice for the latent model is an additive noise model where \( \theta = \tilde{\theta} + w \), with \( w \) drawn from a known noise distribution. Under this model, it can be shown (see Appendix~\ref{app:add_nosie}) that
\begin{equation}
    \nabla_{\tilde{\theta}} \log P(x \mid \tilde{\theta}) 
    = \int \nabla_{\theta} \log P(x \mid \theta) \, P(\theta \mid \tilde{\theta}, x) \, d\theta,
\end{equation}
i.e., the estimated score is a convolution of the true score with a convolution kernel $P(\theta \mid \tilde{\theta}, x)$ that is even more sharply peaked than $P(\theta \mid \tilde{\theta})$. 

Pseudocode for both the decomposable and indecomposable cases is provided in Appendix~\ref{app:alg}.



\section{Experiments}
\label{sec:ex}
To demonstrate the effectiveness of our approach, we present experiments aimed at verifying that the trained Fisher score can be used for Fisher forecasting and Bayesian inference. Details of the network architecture and other relevant parameters can be found in Appendix~\ref{app:network}. The code used for these experiments is available at \url{https://github.com/suicee/FisherScoreMatching}.

\subsection{Toy example}

To demonstrate the application of our method to decomposable models, we first consider a simple linear Gaussian model. The full model is given by \( x \mid \theta \sim \mathcal{N}(\theta, \Sigma_n) \), where \( \theta \in \mathbb{R}^2 \) is the parameter of interest and \( \Sigma_n = \begin{bmatrix} 1 & 0.5 \\ 0.5 & 1 \end{bmatrix} \).

We decompose the model according to Equation~\ref{gaussian_model}, and choose the latent model covariance to be \( \Sigma_{\theta^*} = 0.4 \cdot \mathbf{I} \). We then sample \( \theta \) from a uniform prior over \([-3, 3]^2\), and generate \( \theta^* \) and \( x \) using the conditional distributions defined by the decomposition. We generate 100,000 sample triplets for training.

\begin{figure}[!htb]
    \centering
\includegraphics[width=0.8\linewidth]{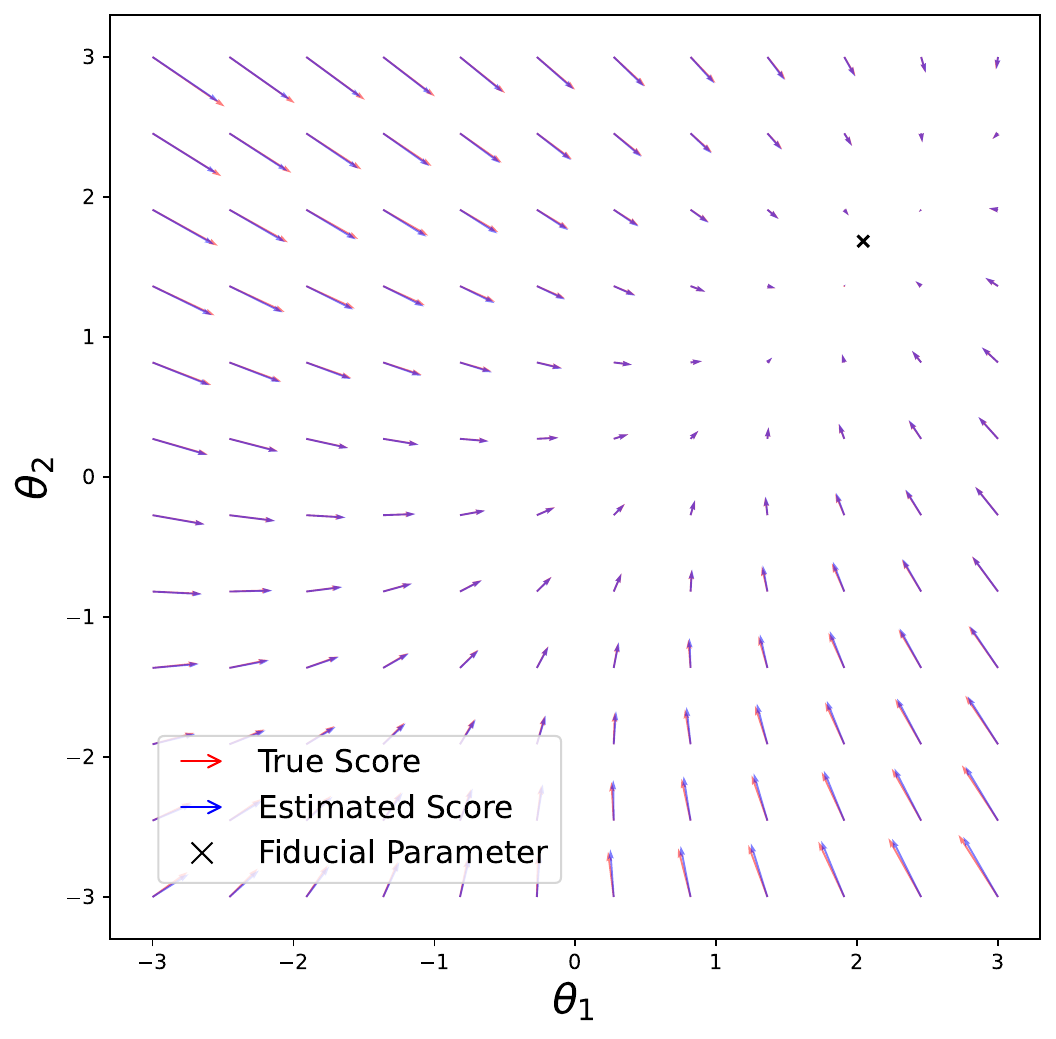}
    \caption{Comparison of the learned Fisher score (blue arrows) and the analytical Fisher score (red arrows) across the parameter space for a fixed observation. The black dot denotes the fiducial parameter value corresponding to the observation. Arrows originate at parameter locations and point in the direction of the Fisher score. The close alignment of the two sets of arrows indicates accurate recovery of the Fisher score.}
    \label{fig:simple_gaussian_vf}
\end{figure}

For evaluation, we sample a random observation \( x_{\text{obs}} \), and compute the Fisher score across the entire parameter space. As shown in Figure~\ref{fig:simple_gaussian_vf}, the learned score field closely matches the analytical Fisher score. This verifies that our method can recover the correct Fisher score using only the score of the latent model.

\subsection{Weak Lensing Example}

As a more realistic demonstration, we apply our method to \jaxcosmo{}\footnote{\url{https://github.com/DifferentiableUniverseInitiative/jax_cosmo}} , a library for automatically differentiable cosmological theory computations using  JAX \cite{jaxcosmo}.  It provides access to gradients of theoretical predictions of various cross-correlation statistics of weak lensing and galaxy observations with respect to cosmological parameters, making it well-suited for validating our approach.

In this experiment, we vary two key cosmological parameters in the standard model of cosmology: the cold dark matter density \(\Omega_c\) and the matter fluctuation amplitude \(\sigma_8\). For each pair \((\Omega_c, \sigma_8)\), we use \jaxcosmo\ to simulate projected two-point galaxy lensing angular power spectra. The angular power spectra \(C_\ell\) are computed at five logarithmically spaced multipole between \(\ell = 10\) and \(\ell = 1000\). All other cosmological parameters and configuration settings are fixed to their default values as defined in the package.

We assume the data likelihood is Gaussian:
\begin{equation}
    x \mid \theta \sim \mathcal{N}(\mu(\theta), \Sigma_{\text{fid}}),
\end{equation}
where \(x = C_{\ell}\), \(\theta = (\Omega_c, \sigma_8)\), \(\mu(\theta)\) is the mean spectrum computed by \jaxcosmo, and \(\Sigma_{\text{fid}}\) is a fixed covariance matrix evaluated at a fiducial cosmology.

Under this Gaussian model and using the differentiability of \jaxcosmo, we can compute the ground truth Fisher score and Fisher information matrix as:
\begin{equation}
\begin{aligned}
    s^*(x, \theta) &= \left( \nabla_{\theta} \mu(\theta) \right)^\top \Sigma_{fid}^{-1} (x - \mu(\theta)), \\
    \mathcal{I^*}(\theta) &= \left( \nabla_{\theta} \mu(\theta) \right)^\top \Sigma_{fid}^{-1} \left( \nabla_{\theta} \mu(\theta) \right).
\end{aligned}
\label{eq:gassian_fisher_score_and_info}
\end{equation}

This model is not obviously decomposable, so we introduce an auxiliary latent variable \(\tilde{\theta}\). We sample \(\tilde{\theta}\) from the  ranges \([0.2, 0.4]\) for \(\Omega_c\) and \([0.70, 0.90]\) for \(\sigma_8\). Then, we generate \(\theta\) by adding Gaussian noise: \(\theta \sim \mathcal{N}(\tilde{\theta}, \sigma^2 \cdot \mathbf{I})\), with \(\sigma = 10^{-3}\). Once \(\theta\) is sampled, we compute \(C_\ell\) using \jaxcosmo. We generate 100,000 sample triplets \((\tilde{\theta}, \theta, x)\) for training.

\begin{figure}[!htb]
    \centering
    \includegraphics[width=0.8\linewidth]{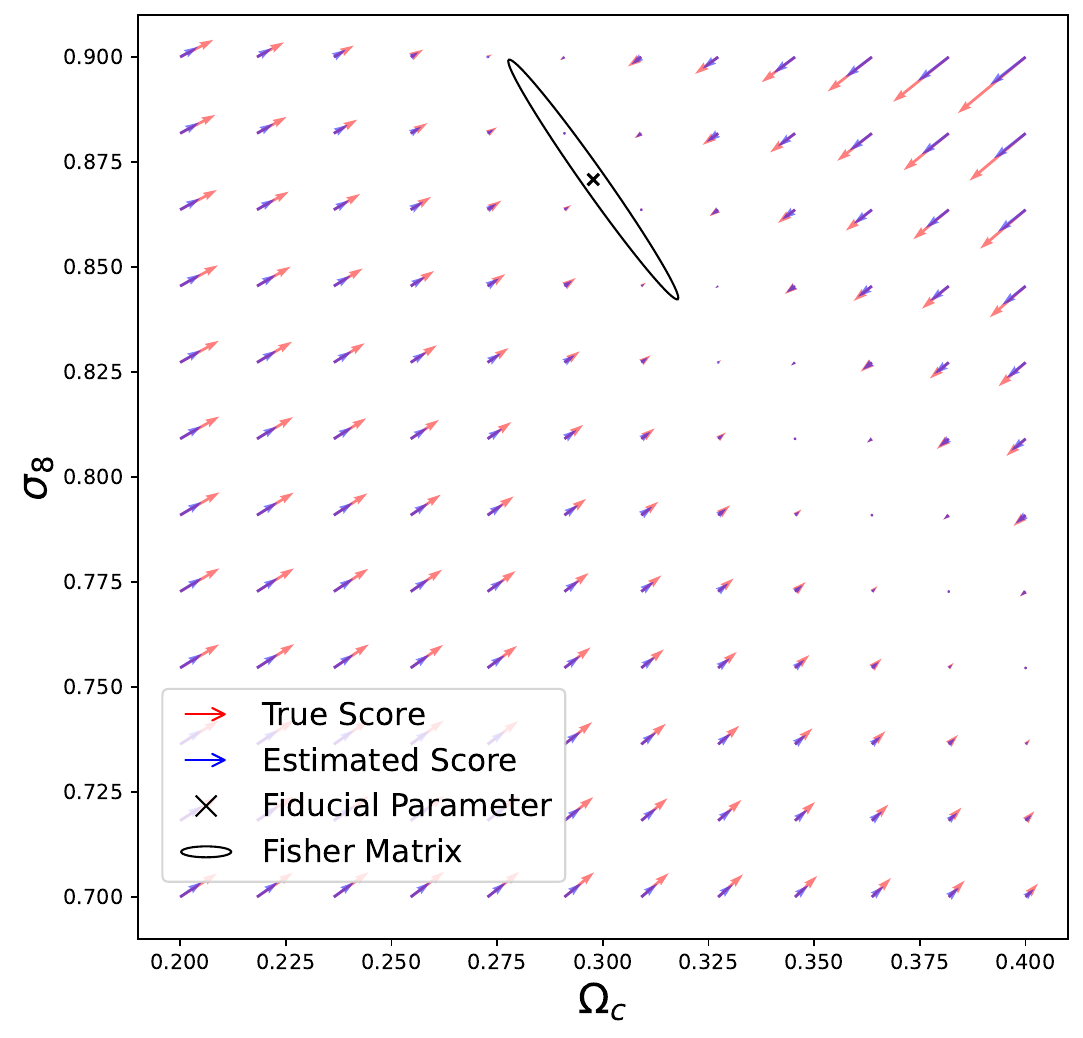}
    \caption{
        Extrapolation of the learned Fisher score (blue arrows) \textit{vs}  the true Fisher score (red arrows) in the weak lensing example, for a fixed observation. The black cross indicates the fiducial parameters and the Fisher matrix contours are shown for scale. For  implausible parameters far away from the fiducial,  the score is underestimated by  $\sim 40$\%, but the direction is still accurately captured. This does not affect sampling of high likelihood regions or maximum likelihood estimation.
    }
    \label{fig:weaklensing_vf}
\end{figure}

As in the toy example, we first evaluate the learned Fisher score across the parameter space for a fixed mock observation. Figure~\ref{fig:weaklensing_vf} shows the learned score field accurately matches the true score directions (obtained via autodifferentiation), which are approximately orthogonal to the degeneracy direction between parameters expected from weak lensing correlation analysis \cite{wl_degeneracy}. 

The training set will contain few relevant examples for  parameter values that have small probability to have generated the mock observation. This leads to significant errors in the magnitude of the estimated score vectors for parameters that are implausible given the data. In spite of that, the model stably extrapolates across the entire training range to within a factor of 2.  This will not affect  applications such as forecasting and inference where the model encounters plausible combinations of data and parameters, as evidenced in the following experiments.

\begin{figure}[!htb]
    \centering
    \includegraphics[width=0.8\linewidth]{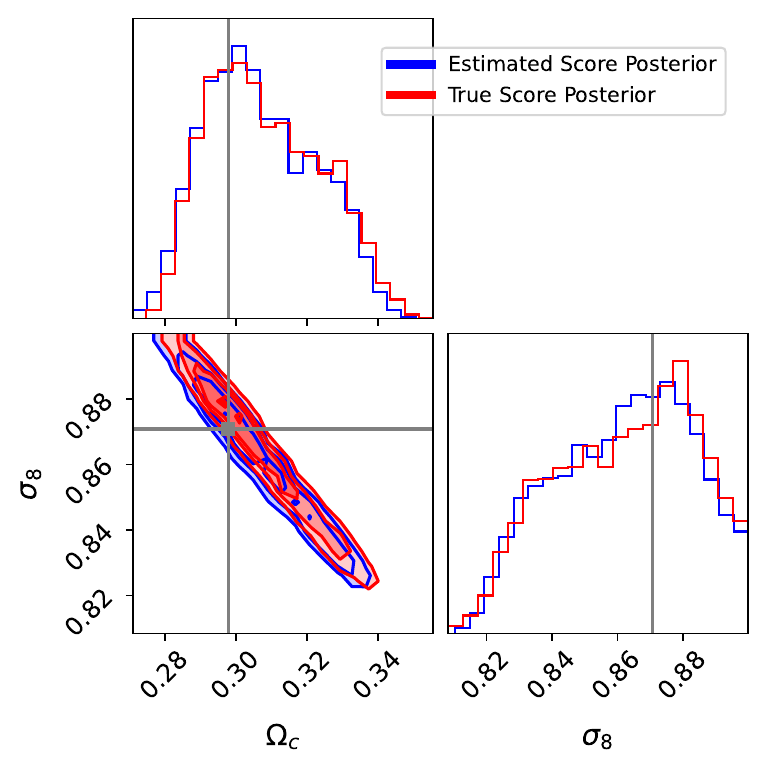}
    \caption{
        Posterior distribution obtained via HMC using the learned Fisher scores (blue contours) versus the true Fisher scores (red contours). The gray cross marks the fiducial parameter used to generate the mock observation.
    }
    \label{fig:enter-label}
\end{figure}

By supplying a prior (uniform in the ranges \([0.2, 0.4]\) for \(\Omega_c\) and \([0.70, 0.90]\) for \(\sigma_8\)) we use HMC to sample the posterior based on the estimated score vector field. As shown in Figure~\ref{fig:enter-label}, the posterior contours from the estimated scores (blue) align closely with those from the true scores (red), confirming that the learned scores are suitable for Bayesian inference.

\begin{figure}[!htb]
    \centering
    \includegraphics[width=0.8\linewidth]{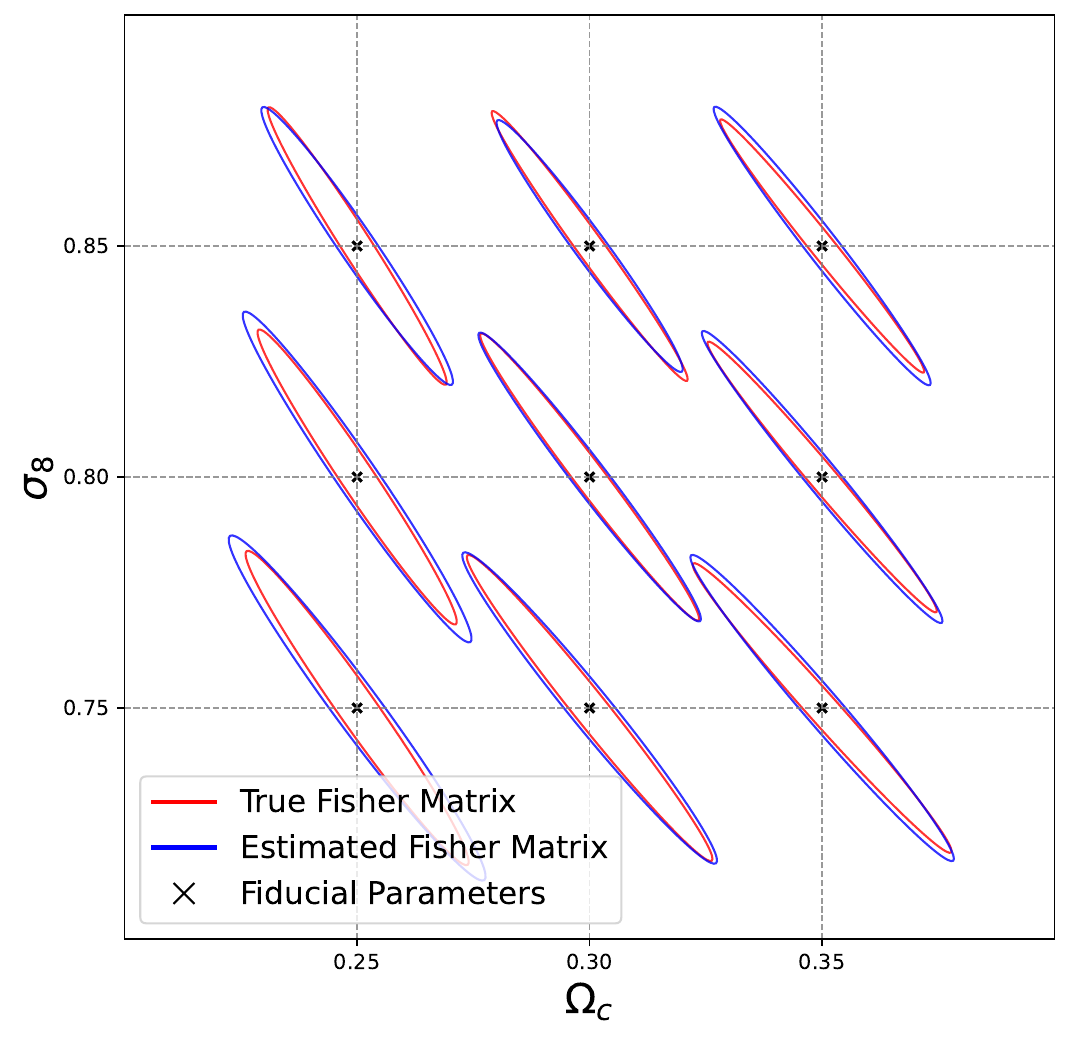}
    \caption{
        Comparison of estimated Fisher matrices (blue contours) and true Fisher matrices (red contours). Each contour corresponds to a Fisher matrix estimated at its center (indicated by a cross).
    }
    \label{fig:weaklensing_fim}
\end{figure}

As a second example, we can  estimate the Fisher information matrix at any point in parameter space $\theta_0$ by computing the covariance of the score model 
\begin{equation}
    \mathcal{I}(\theta_0) = E_{p(x \mid \theta_0)}\left[ s(x, \theta_0) s(x, \theta_0)^\top \right],
\end{equation}
where the expectation is approximated using samples drawn from \(p(x \mid \theta_0)\). We evaluate the Fisher matrix on a grid of parameter values and compare the results to those obtained from Equation~\ref{eq:gassian_fisher_score_and_info}. As shown in Figure~\ref{fig:weaklensing_fim}, the estimated contours align closely with the true Fisher matrices.

\section{Conclusion}
\label{sec:conclusion}
We propose a score-matching-based method to estimate the Fisher score only from simulations by introducing a latent model to generate the parameters of the target model.

Through a toy example and a cosmological application, we show that this approach performs well, both in scenarios where a natural latent is suggested by the model structure (decomposable), or when it is added as a perturbation (non-decomposable).  Once trained, the model can estimate the Fisher score at any given parameter-data pair. This enables gradient-based Bayesian inference (e.g., via HMC) and fast Fisher forecasting, as demonstrated in our experiments.

Modeling the Fisher score function differs conceptually and in practice from training a conditional diffusion model by separating the act of choosing the prior from the construction of the training set. When training a diffusion model the choice of training set determines the prior, whereas the Fisher score approximation that minimizes our loss function, Equation~\ref{eq:FisherScoreLoss}, is independent of the prior. In practice, successful training, of course, requires a training set of sufficient variety and size.

Several directions remain for future work. On the methodological side,  large training datasets are typically required for an accurate score estimates in low-probability regions; finding strategies to reduce this would be valuable. 
On the application side, it would be interesting to apply this framework to high-dimensional problems, such as field-level Fisher analysis or simulation-based inference, where existing techniques face significant challenges.

\bibliography{ref}
\bibliographystyle{icml2025}

\newpage
\appendix
\onecolumn
\section{Bayesian Least Squares Estimation}
\label{app:bls}
Assume \((x, y) \sim P(x, y)\), and let \(g(x)\) be a target function. To estimate \(g(x)\) from observations \(y\), we consider minimizing the mean squared error:
\[
\mathcal{L}(f) = \mathbb{E}_{P(x, y)} \left[ (f(y) - g(x))^2 \right].
\]
The minimizer of this loss is the Bayes least squares estimator:
\[
f^*(y) = \mathbb{E}_{P(x \mid y)} [g(x)].
\]
This follows from the decomposition:
\[
\begin{aligned}
\mathcal{L}(f)
&= \mathbb{E}_{P(y)} \left[ \mathbb{E}_{P(x \mid y)} \left[ (f(y) - g(x))^2 \right] \right] \\
&= \mathbb{E}_{P(y)} \left[ \left(f(y) - \mathbb{E}_{P(x \mid y)}[g(x)] \right)^2 + \operatorname{Var}_{P(x \mid y)}[g(x)] \right],
\end{aligned}
\]
where \(\operatorname{Var}_{P(x \mid y)}[g(x)] = \mathbb{E}_{P(x \mid y)}\left[(g(x) - \mathbb{E}_{P(x \mid y)}[g(x)])^2\right]\) is independent of \(f\). Thus, the optimal estimator matches the conditional mean.

\section{Error Analysis for Indecomposable Models with Additive Noise}
\label{app:add_nosie}
When a natural latent variable is unavailable, we introduce an auxiliary variable \(\tilde{\theta}\) and define a latent model \(P(\theta \mid \tilde{\theta})\). In this case, our method estimates the Fisher score of the extended model, \(\nabla_{\tilde{\theta}} \log P(x \mid \tilde{\theta})\), rather than the original model.

Assume an additive noise model:
\[
\theta = \tilde{\theta} + w, \quad \text{where } w \sim p(w).
\]
Then the marginal likelihood under the extended model is:
\[
P(x \mid \tilde{\theta}) = \int P(x \mid \theta) \, p(\theta \mid \tilde{\theta}) \, d\theta = \int P(x \mid \tilde{\theta} + w) \, p(w) \, dw.
\]
Taking the gradient with respect to \(\tilde{\theta}\), we have:
\[
\begin{aligned}
\nabla_{\tilde{\theta}} P(x \mid \tilde{\theta}) 
&= \int \nabla_{\tilde{\theta}} P(x \mid \tilde{\theta} + w) \, p(w) \, dw \\
&= \int \nabla_{\tilde{\theta}} \log P(x \mid \tilde{\theta} + w) \, P(x \mid \tilde{\theta} + w) \, p(w) \, dw.
\end{aligned}
\]
Using the definition of the Fisher score, we obtain:
\[
\nabla_{\tilde{\theta}} \log P(x \mid \tilde{\theta}) 
= \frac{\nabla_{\tilde{\theta}} P(x \mid \tilde{\theta})}{P(x \mid \tilde{\theta})}
= \int \nabla_{\tilde{\theta}} \log P(x \mid \tilde{\theta} + w) \, \frac{P(x \mid \tilde{\theta} + w) \, p(w)}{P(x \mid \tilde{\theta})} \, dw.
\]
Substituting \(\theta = \tilde{\theta} + w\), and applying Bayes' theorem, we rewrite the integrand as 
\[
\nabla_{\tilde{\theta}} \log P(x \mid \tilde{\theta}) 
= \int \nabla_{\theta} \log P(x \mid \theta) \, P(\theta \mid \tilde{\theta}, x) \, d\theta.
\]

This result shows that the estimated score is a smoothed version of the true score \(\nabla_{\theta} \log P(x \mid \theta)\). The accuracy of this approximation depends on the sharpness of \(P(\theta \mid \tilde{\theta})\), which bounds the sharpness of $P(\theta \mid \tilde{\theta}, x) $. When it is highly concentrated (e.g., close to a delta function), the smoothed score approaches the true Fisher score.

\section{Fisher Score Matching Algorithms}
\label{app:alg}
\noindent
\begin{minipage}[t]{0.48\linewidth}
\begin{algorithm}[H]
   \caption{FSM for Decomposable Model}
   \label{alg:decomposable}
\begin{algorithmic}
   \STATE {\bfseries Input:} Simulator $x \mid \theta \sim \int p(x \mid \theta^*)p(\theta^* \mid \theta)d\theta^*$, model $s(x, \theta)$, proposal $p(\theta)$
   \WHILE{not converged}
       \STATE Sample batch: $(\theta, \theta^*, x) \sim p(\theta)p(\theta^* \mid \theta)p(x \mid \theta^*)$
       \STATE Compute target: $\nabla_{\theta} \log p(\theta^* \mid \theta)$
       \STATE Predict score: $\hat{s} = s(x, \theta)$
       \STATE Compute loss: $L = \|\hat{s} - \text{target}\|^2$
       \STATE Update model parameters using $\nabla L$
   \ENDWHILE
   \STATE {\bfseries Output:} Trained estimator $s(x, \theta)\approx \nabla_{\theta} \log p(x \mid \theta)$
\end{algorithmic}
\end{algorithm}
\end{minipage}
\hfill
\begin{minipage}[t]{0.48\linewidth}
\begin{algorithm}[H]
   \caption{FSM for Indecomposable Model}
   \label{alg:indecomposable}
\begin{algorithmic}
   \STATE {\bfseries Input:} Simulator $x \mid \theta \sim p(x \mid \theta)$, model $s(x, \tilde{\theta})$, proposal $p(\tilde{\theta})$, conditional $p(\theta \mid \tilde{\theta})$
   \WHILE{not converged}
       \STATE Sample batch: $(\tilde{\theta}, \theta, x) \sim p(\tilde{\theta})p(\theta \mid \tilde{\theta})p(x \mid \theta)$
       \STATE Compute target: $\nabla_{\tilde{\theta}} \log p(\theta \mid \tilde{\theta})$
       \STATE Predict score: $\hat{s} = s(x, \tilde{\theta})$
       \STATE Compute loss: $L = \|\hat{s} - \text{target}\|^2$
       \STATE Update model parameters using $\nabla L$
   \ENDWHILE
   \STATE {\bfseries Output:} Trained estimator $s(x, \tilde{\theta})\approx \nabla_{\theta} \log p(x \mid \tilde{\theta})$
\end{algorithmic}
\end{algorithm}
\end{minipage}

\section{Network Details}
\label{app:network}
For the Gaussian model, we use a multilayer perceptron (MLP) with two hidden layers of 64 units each, followed by ELU activation. The network is trained with a learning rate of 1e-2 for 1000 epochs. For the Weak Lensing example, we use an MLP with hidden layers of sizes 64, 128, and 128, also followed by ELU activation. The network is trained with a learning rate of 1e-3 for 10,000 epochs.


\end{document}